# MECHANICAL AND ELECTRICAL NOISE IN THE PVLAS EXPERIMENT

M. Bregant[1], G. Cantatore[1], S. Carusotto[2], G. Di Domenico[3], F. Della Valle[1], U. Gastaldi[4], E. Milotti[5], G. Petrucci[6], E. Polacco[2], G. Ruoso[4], E. Zavattini[1], G. Zavattini[3]

[1] Universita` di Trieste and INFN-Sezione di Trieste, Italy
[2] Universita` di Pisa and INFN-Sezione di Pisa, Italy
[3] Universita` di Ferrara and INFN-Sezione di Ferrara, Italy
[4] INFN-Laboratori Nazionali di Legnaro, Italy
[5] Universita` di Udine and INFN-Sezione di Trieste, Italy
[6] CERN, Geneva, Switzerland

**ABSTRACT.** PVLAS is an experiment which aims at the direct detection of photon-field scattering: it employs optical methods and a large rotating superconducting magnet, and its large, compact structure is affected by both mechanical and electrical noises. This paper introduces briefly the data analysis methods used in the experiment and summarizes the mechanical and electrical noise situation.

## INTRODUCTION

Almost seventy years ago, Heisenberg, Euler and Weisskopf [1,2] predicted that photons may be scattered by a static electric or magnetic field. This is a straighforward prediction of Quantum Electrodynamics [3], and it is believed that such scattering processes, and the associated photon splitting processes, play an important role in astrophysical environments, such as on the surfaces of pulsars and magnetars, or in the weak but very extended galactic magnetic fields, but until now this has never been directly verified in the laboratory. The PVLAS experiment aims at the direct detection of photon-field scattering, and in addition it may also be used to detect low-mass pseudoscalar or scalar particles with two-photon interactions, like the DFSZ axions (see [4] for the theory and [5] for an experimental review). PVLAS employs optical methods and a large rotating superconducting magnet [6]: infrared photons are injected into a Fabry-Perot interferometer and as they bounce back and forth in the magnetic field region between the interferometer mirrors, they may occasionally interact with the magnetic field. The interaction is polarization-dependent and the vacuum between the interferometer mirrors behaves as a uniaxial birefringent medium. The light exiting the interferometer is analyzed by a polarizing prism and a tiny signal is picked up by a low noise photodiode. The large, compact structure of the PVLAS experiment is affected by both mechanical and electrical noises. This paper introduces briefly the data analysis methods employed in the experiment and summarizes the mechanical and electrical noise situation.

## THE QED VACUUM

It is well known that the vacuum of Quantum Field Theory (QFT) is seething with virtual particle pairs that spring into existence and annihilate each other all the time. The charged particle pairs may be viewed either as electric or as magnetic dipoles, and the QED vacuum acts as a polarizable medium. Since photons moving in this QED vacuum interact with the virtual particle pairs, the photon propagator is modified by the presence of the particle pairs, and photons may be scattered by other photons or by an external field.

The interaction Lagrangian was first introduced by Euler and Heisenberg [1] and by Weisskopf [2], and it has been formalized in the context of modern QED by Schwinger [7], who calculated the following effective interaction Lagrangian

$$\mathcal{L} = -\frac{F}{4\pi} - \frac{1}{8\pi^2} \int_0^\infty \frac{ds}{s^2} \exp(-m^2 s)\left\{(es)^2 L - 1 - \frac{2}{3}(es)^2 F\right\}, \tag{1}$$

where $m$ is the electron mass, $e$ is the elementary charge (and $\hbar = c = 1$) and $L$, $F$ and $G$ are defined by the expressions

$$L = iG\frac{\cosh\{es[2(F+iG)]^{1/2}\} + \cosh\{es[2(F-iG)]^{1/2}\}}{\cosh\{es[2(F+iG)]^{1/2}\} - \cosh\{es[2(F-iG)]^{1/2}\}}, \tag{2}$$

and

$$F = \frac{1}{2}(\mathbf{B}^2 - \mathbf{E}^2); \quad G = \mathbf{B}\cdot\mathbf{E}. \tag{3}$$

(the equivalent diagrammatic perturbative expansion is shown in Figure 1). The scalar term containing $F$ in expression (1) is just the classical Lagrangian of the EM field, but there is a novel pseudoscalar term $G$ as well: this means that when a magnetic field is present, the QED vacuum is characterized by the direction of the magnetic field, and it becomes birefringent. This birefringence is actually quite small: in a dipolar field $\mathbf{B}$ the QED vacuum becomes a uniaxial birefringent medium with a difference between the refractive indexes

$$\Delta n = |n_\parallel - n_\perp| \approx \left(4\cdot 10^{-24}\ \mathrm{Tesla}^{-2}\right) B_0^2 \sin 2\theta, \tag{4}$$

($\theta$ is the angle between the external magnetic field and the electric field of the linearly polarized light) so that linearly polarized light propagating in this medium acquires a small ellipticity

$$\Psi = \frac{\pi}{\lambda} L \Delta n \approx \left(12.5\cdot 10^{-24}\ \mathrm{Tesla}^{-2}\right)\left(\frac{L}{\lambda}\right) B_0^2 \sin 2\theta \tag{5}$$

($L$ is the path length inside the birefringent medium, and $\lambda$ is the wavelength of light) and thus, even with the intense fields achievable with the present superconducting magnet technology, the difference (4) remains exceedingly small, and for this reason the light-field interaction has long remained undetected.

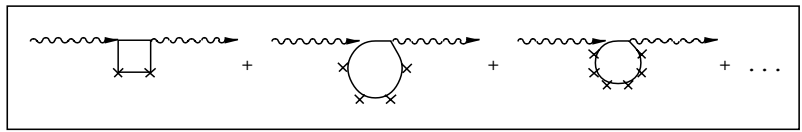

**FIGURE 1.** The diagrammatic perturbative expansion equivalent to the EHW result and to the Schwinger Lagrangian (1) (the crosses denote interactions with the external field and the permutations of the vertices are implicitly assumed).

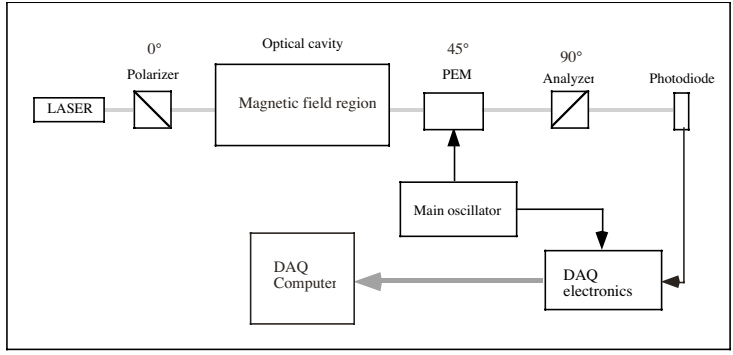

**FIGURE 2.** Schematic layout of the PVLAS experiment. Infrared light is emitted by an infrared Nd-YAG laser, and it is polarized by a Glan prism. The laser beam travels through a magnetic field region where it picks up a small ellipticity, then through a photoelastic modulator (PEM). or through a stress optic modulator (SOM) which introduces an additional ellipticity modulation at high frequency, and finally through another Glan prism which acts as a polarization analyzer, before being detected by the a low-noise photodiode.

## THE PVLAS EXPERIMENT

The PVLAS experiment is a second-generation experiment, which follows a series of preliminary tests carried out at CERN in the early 1980's [6] and the experiment LAS at Brookhaven [5,8]; it is located in Legnaro, near Padua (Italy) in one of the laboratories of the italian National Institute for Nuclear Physics (INFN) [9]. The experimental apparatus is basically a very sensitive ellipsometer which should detect the tiny ellipticity (5) picked up by a beam of linearly polarized light as it travels through an intense magnetic (dipole) field, and it is shown schematically in Figure 2. The effect is modulated by rotating the dipole magnet at about 0.3 rotations/s, so that the ellipticity is

$$\Psi = \Psi_0 \sin 2\omega_M t = \left(12.5 \cdot 10^{-24} \text{ Tesla}^{-2}\right)\left(\frac{L}{\lambda}\right) B_0^2 \sin 2\omega_M t \qquad (6)$$

where $\omega_M \approx 2\pi \cdot (0.3\,\text{Hz}) \approx 2$ radians·s$^{-1}$. This is a very low frequency and any signal at such a low frequency is usually drowned in the *1/f* noise background. Moreover the final photodiode is a square-law detector and one would thus detect an even smaller signal ($\Psi^2$ instead of $\Psi$). For these reasons an optical modulator introduces an additional, controlled, ellipticity at much higher frequency (we have used both a photoelastic modulator [10] at about 20 KHz and a stress-elastic modulator [11] at about 500 Hz): this has the combined advantage of shifting the sought-after effect near the modulator frequency and of linearizing the photodiode response. The main Fourier components in the photodiode signal are listed in Table 1, where $\Gamma$ is the static birefringence introduced by the optical components, $\theta_M$ is the phase of the magnetic field (with respect to a time marker) and $\eta_0 \cos(\omega_{PEM} t + \theta_{PEM})$ is the ellipticity introduced by the modulator. Since the ellipticity signal depends linearly on the path-length *L* and quadratically on the field intensity *B*, it is very important to achieve the highest possible field and the longest possible path-length: we use a one-meter-long superconducting dipole magnet designed and tested to reach 8.8 T (though we use it only up to 6.6 T), and we increase the path-length with a Fabry-Perot resonator. The superconducting magnet is wound with Nb-Ti wire and requires LHe cooling and therefore a large cryostat which must rotate with the magnet: a schematic layout of the cryostat is shown in Figure 3. Figure 4 shows the magnet when it was lowered in the cryostat.

**TABLE 1.** The main Fourier components in the photodiode signal.

| angular frequency | normalized power | phase |
|---|---|---|
| 0 | $\sigma^2 + \Gamma^2 + \eta_0^2/2$ | |
| $\omega_M$ | $2\Gamma\Psi_0$ | $\theta_M$ |
| $2\omega_M$ | $\Psi_0^2/2$ | $2\theta_M$ |
| $\omega_{PEM}$ | $2\Gamma\eta_0$ | $\theta_{PEM}$ |
| $\omega_{PEM} \pm \omega_M$ | $\eta_0\Psi_0$ | $\theta_{PEM} \pm \theta_M$ |
| $2\omega_{PEM}$ | $\eta_0^2/2$ | $2\theta_{PEM}$ |

The Fabry-Perot resonator is more than 6 m long, and is thus longer than the cryostat: with it we have reached a very high finesse ($\sim 10^5$) and a record Q-value ($\sim 10^{12}$) (achieved with an accurate stabilization of the Nd-YAG LASER frequency by means of the Pound-Drever technique [12], see ref. [13] for a discussion): a complete optical layout is shown in Figure 5, and the overall mechanical layout with the rotating cryostat and the optical tables is shown in Figure 6. Figures 7 and 8 show the support structures before and after completion.

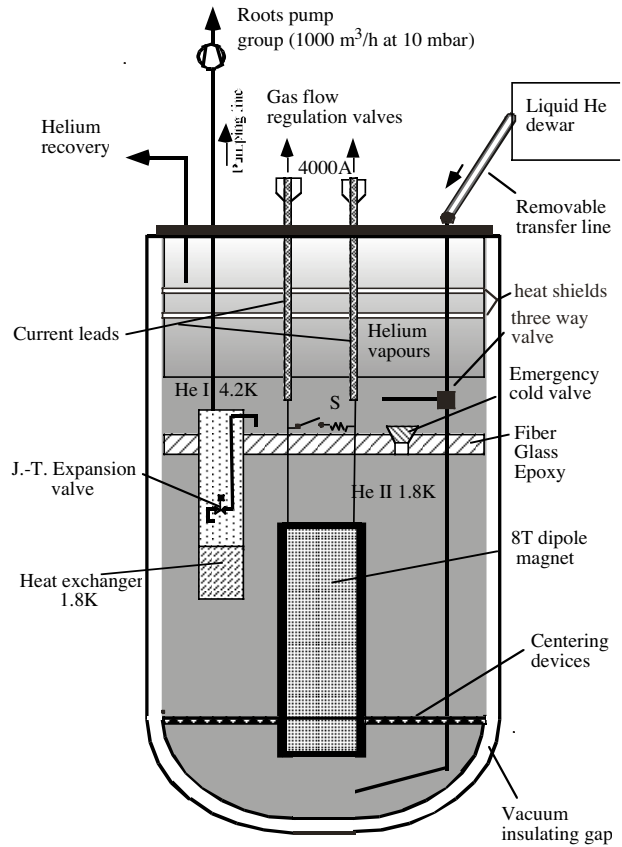

**FIGURE 3.** Schematic layout of the PVLAS cryostat.

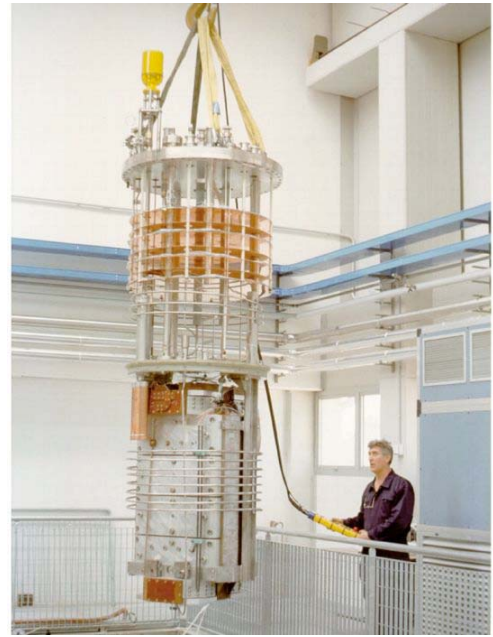

**FIGURE 4.** The magnet and the upper heat-shields being lowered in the cryostat.

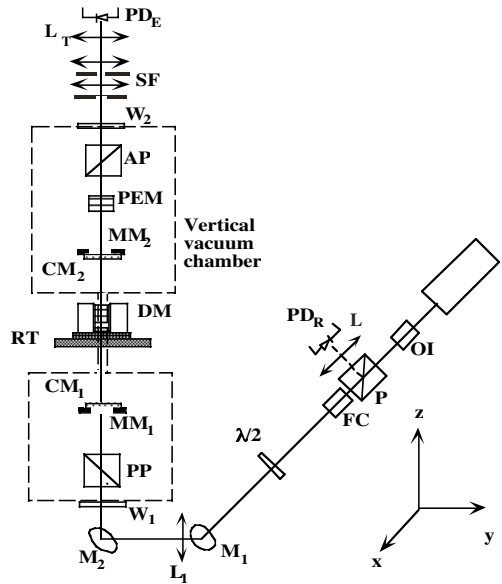

**FIGURE 5.** Detailed optical layout. Legend: OI: Optical Isolator; P,PP,AP Polarizer prisms: $L_R,L_T$:Lenses; $PD_R,PD_E$ Photodiodes; FC: Faraday Cell; $\lambda/2$ : Half wave plate; $L_1$: Matching lens; $M_1,M_2$: steering mirrors $W_1,W_2$: Windows $MM_1,MM_2$ Tilting stages; $CM_1,CM_2$: Cavity mirrors: RT: Rotating Table; DM: Dipole magnet; PEM PhotoElastic Modulator; SF: Spatial Filter.

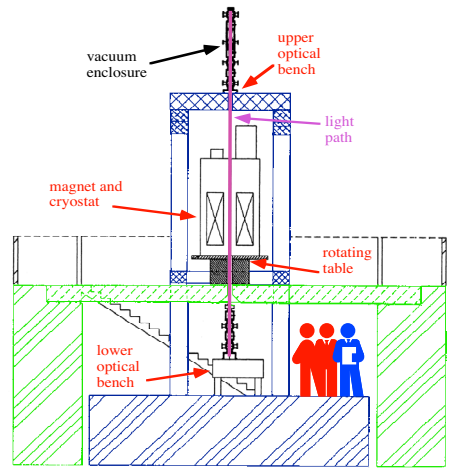

**FIGURE 6.** The overall mechanical layout. The experimental apparatus sits in a hole dug in an experimental hall in Legnaro; the vertical structure that supports the optics (lower optical table plus Fabry-Perot cavity plus upper optical table) is made of granite, and it rests on an isolation raft at the bottom of the hole; the raft is a single block of reinforced concrete supported by the underlying sand bed, and mechanically decoupled from the outer walls of the hole. The rotating cryostat that holds the magnet is also decoupled from the support structure and it sits on a concrete bridge that joins two sides of the hole.

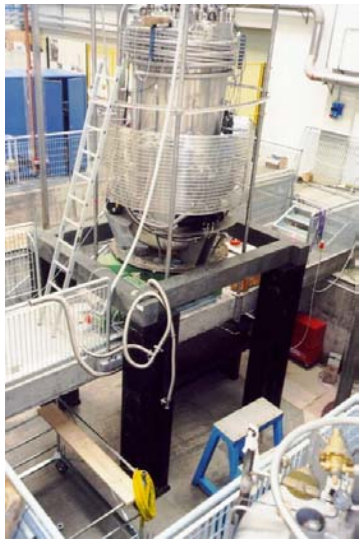

**FIGURE 7.** The experimental apparatus during construction: the rotating table is clearly visible under the cryostat, and only the lower part of granite structure is complete, the lower and the upper optical tables are still missing.

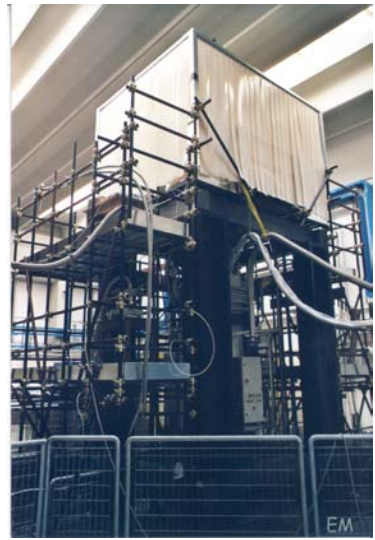

**FIGURE 8.** The experimental apparatus after completion (August 2000): the cryostat is nearly invisible behind the granite columns that support the upper optical table. The table itself is screened by a plastic hut that protects it from dirt and environmental noise. The outer structure made of light tubing is mechanically decoupled from the granite columns and is needed to access the apparatus.

## NOISE SOURCES AND SYSTEMATIC EFFECTS

As we have seen in the previous section the heart of the apparatus is a high-Q Fabry-Perot resonator: the resonator is 6.4 m long and is partially contained in long quartz tube with a internal diameter of about 24 mm. This tube resides in an axial hole in the cryostat, and both the tube and the mirror may be subjected to environmental vibrations, both seismic and acoustical. In addition, the ellipticity modulator may also pick up enviromental noise (an unstable beam position on the modulator translates into an additional random ellipticity modulation because of nonuniformities of the modulator substrate). Several sources may contribute to a diffuse noise background: the vacuum pumps, the rotating table, cars on a nearby road, etc. Many of these sources have a fairly well defined operating frequency, but may be spread on a comparatively large spectral region because of nonlinear couplings. Random disturbances with a flat or nearly flat spectrum are undesirable because they drown in background noise the Fourier components given in table 1, especially the information-carrying sidebands, but systematic effects are even more dangerous, because they produce fake spectral peaks which are not easy to disentagle from the real physical effect. Apart from the obvious and easily controlled electrical cross-talk signals, there are unfortunately several such systematic effects due to the coupling between the rotating magnetic field and many optical, electrical and mechanical components in the apparatus. A very partial list of systematics includes:

1. mechanical effects
    - generic mirror motion (due to parasitic mechanical coupling to rotating table)
    - piezoelectricity in conductors (deformed by acoustical or mechanical effects)
2. magneto-mechanical effects
    - ferromagnetic coupling with supporting structures
    - magnetostriction in supporting structures
3. electromagnetic effects
    - direct induction in conductors
    - direct induction in acquisition electronics
    - Faraday and Cotton-Mouton effects in gases or glass

## DATA ACQUISITION AND ANALYSIS

Data are collected from several sensors using 16-bit (and in the near future 24-bit) ADC's, and special care is devoted to the main photodiode that detects the light signal from the cavity. The acquisition is performed at low frequency (about 10 Hz) in a self-triggered mode (tick marks on the rotating table trigger the acquisition apparatus and allow a perfect synchronization with the the physical effect) and at high frequency (8200 Hz, but may be as low as 4100 Hz or as fast as 90 KHz) in a free-running mode (an internal clock in the acquisition computer triggers data acquisition at a frequency above the Nyquist frequency for the ellipticity modulator): all the signals are written to disk for offline analysis. Online monitoring is provided by a few spectrum analyzers. The low frequency signals also require demodulating lock-in amplifiers to remove the modulator frequency.
Figure 9 shows an example of a real physical signal (Cotton-Mouton effect in nitrogen, see for instance ref [14]) read out with the low-frequency data acquisition, and Figure 10 shows the corresponding power spectral density. Figures 11 and 12 show the power spectral density obtained with the high-frequency data acquisition system: a comparison between the low-frequency and the high-frequency data may eventually help to detect the sources of random noise and systematic effects.

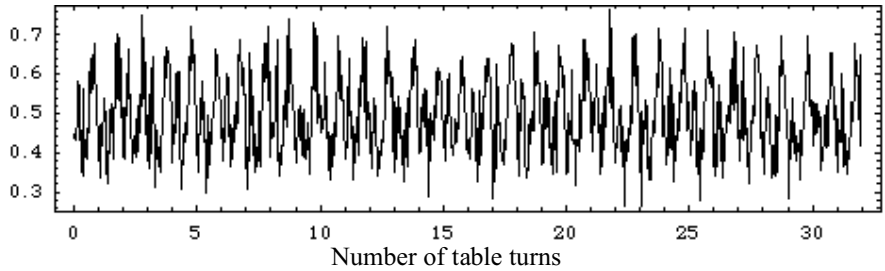

**FIGURE 9.** Initial portion of the main signal taken in run 353 (november 2001). In this case the cavity was not empty but contained nitrogen with a pressure of 0.0052 mbar, and the magnet current was $I = 859$ A, corresponding to a field intensity $B = 3.28$ T. The horizontal scale is in units of table turns, while the vertical scale is in Volt at the ADC input (the combined transimpedance amplification factor before the ADC is $10^{10}$ V/A so that the photocurrent is of the order of $10^{-11}$-$10^{-10}$ A). The signal shown here displays a real physical effect, the Cotton-Mouton effect in nitrogen, which is a few orders of magnitude greater than the QED effect in vacuum, and is used to calibrate and check the apparatus.

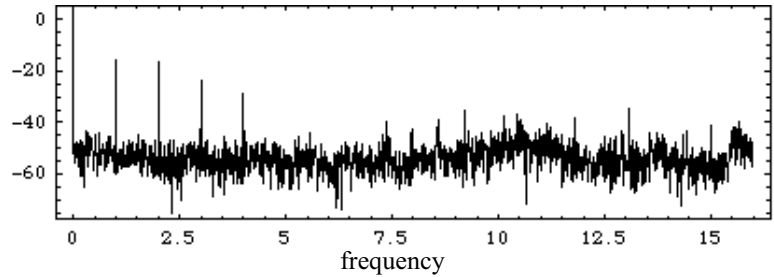

**FIGURE 10.** Power spectral density of the signal shown in Figure 9: amplitude (dBVrms) vs. frequency (turns$^{-1}$). Four peaks emerge from a mostly flat background, corresponding to the first, second, third and fourth harmonic of the magnet rotation frequency: at present we can only extract physical information from the second harmonic.

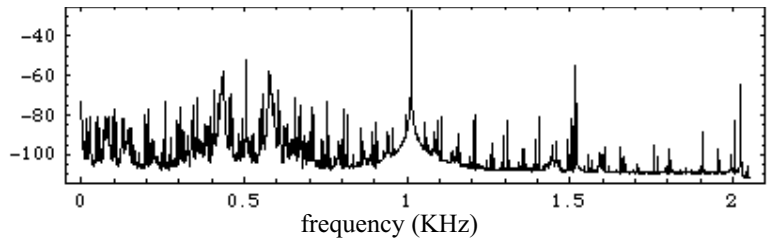

**FIGURE 11.** Power spectral density of the photodiode signal obtained with the high frequency data acquisition ADC (modulator frequency = 506 Hz, sampling frequency = 4100 Hz, transimpedance amplification = $10^7$ V/A): amplitude (dBVrms) vs. frequency (Hz). In addition to some of the Fourier components listed in table 1 there are many other peaks outside the band of interest. This is an overall view and the interesting frequency range is compressed in a single channel around 0.5 KHz.

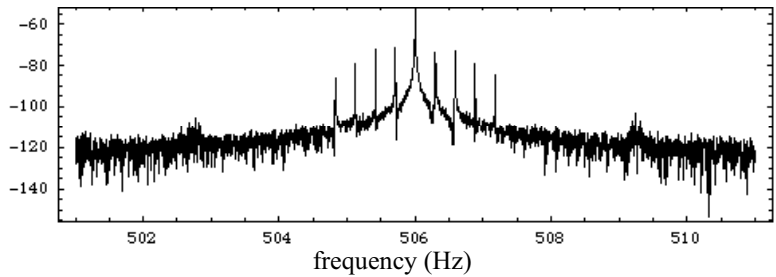

**FIGURE 12.** Expanded view of the power spectral density near the stress-optic modulator frequency peak at 506 Hz. This spectrum has been obtained with $2^{20}$ samples, and the resolution is about 4 mHz. The four sidebands visible in Figure 10 stand out very clearly near the peak of the modulator signal. The sidebands are not exactly symmetrical about the modulator peak: this may indicate an unwanted phase modulation from some still unidentified magnetomechanical effect.

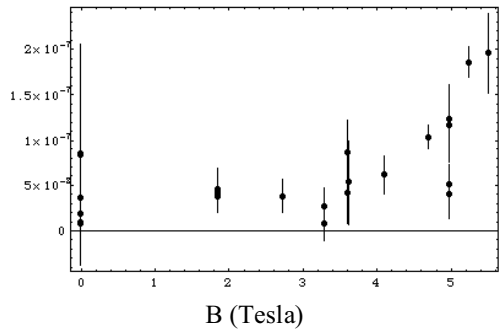

**FIGURE 13.** Measured ellipticity vs. B (Tesla) in vacuum. Notice the linear growth after 3 T: we believe that as the saturation value is reached, the iron cannot contain the field in the iron yoke, and that this influences some electrical or optical element in the apparatus. This hypothesis is further supported by numerical magnetic field calculations. We have not yet identified the sensitive elements.

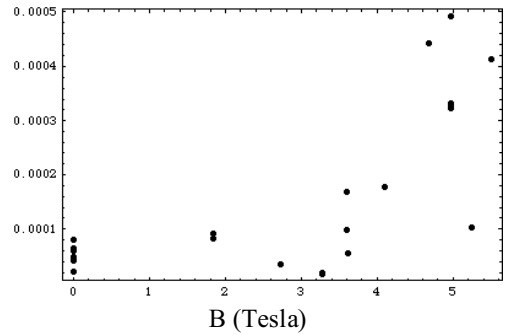

**FIGURE 14.** Measured LASER beam displacement ($mm_{RMS}$) vs B (Tesla) at twice the magnet rotation frequency (this is the frequency at which we measure ellipticity). Notice that the RMS beam displacement behaves like the ellipticity in fig. 13: this provides further support to the hypothesis discussed in Figure 13.

## FUTURE PERSPECTIVES

Like all high-precision experiments, PVLAS is very challenging, and must solve problems due to systematic effects and noise in order to measure the minute QED birefringence. Now we believe that we are on the verge of understanding how the magnetic field couples with our apparatus: Figures 13 and 14 show that the systematic signal that we observe is probably due to the influence of the magnetic field that escapes from the magnet yoke when the iron saturates. This means that a proper screening of the magnetic field should cure the apparatus and improve both our measurement sensibility and remove most systematics. If this is correct, then we shall soon be able not only to give tight limits to the axion mass [4,5], and measure accurately the Cotton-Mouton constants of several gases at low pressure [14], but also to observe the long-sought QED effect.